\documentclass[a4paper]{article}
\usepackage{color,graphicx}
\usepackage{amsmath}
\usepackage{amsthm}
\usepackage{amssymb}
\usepackage{amsfonts}
\RequirePackage{ifpdf} 
\ifpdf
\usepackage[pdftex,bookmarks=true,
		   bookmarksnumbered=false,
		   bookmarksopen=false,
		   colorlinks=true,
		   linkcolor=webred] {hyperref}
\definecolor{webgreen}{rgb}{0, 0.5, 0} 
\definecolor{webblue}{rgb}{0, 0, 0.5} 
\definecolor{webred}{rgb}{0.5, 0, 0}   
\else
\newcommand{\href}[2]{ #1 }
\fi
\title{Netsukuku topology}
\author{http://netsukuku.freaknet.org\\AlpT (@freaknet.org)}
\begin{document}
\maketitle
\begin{abstract}
	In this document, we describe the fractal structure of the Netsukuku
	topology. Moreover, we show how it is possible to use the QSPN v2 on
	the high levels of the fractal.
\end{abstract}
\pagenumbering{roman}
\pagebreak
\begin{small}
  This document is part of Netsukuku.\\
  Copyright \copyright 2007 Andrea Lo Pumo aka AlpT $<$alpt@freaknet.org$>$.
  All rights reserved.

  This document is free; you can redistribute it and/or modify it
  under the terms of the GNU General Public License as published by
  the Free Software Foundation; either version 2 of the License, or
  (at your option) any later version.

  This document is distributed in the hope that it will be useful, but
  WITHOUT ANY WARRANTY; without even the implied warranty of
  MERCHANTABILITY or FITNESS FOR A PARTICULAR PURPOSE\@.  See the GNU
  General Public License for more details.

  You should have received a copy of the GNU General Public License
  along with this document; if not, write to the Free Software
  Foundation, Inc., 675 Mass Ave, Cambridge, MA 02139, USA.
\end{small}

\clearpage
\tableofcontents
\clearpage
\pagenumbering{arabic}

\section{Preface}
\label{sec:preface}

We're assuming that you already know the basics of the QSPN. If not, read the
QSPN document first: \cite{qspndoc}.

\section{The general idea}
\label{sec:general_idea}

The aim of Netsukuku is to be a (physical) scalable mesh network, completely
distributed and decentralised, anonymous and autonomous.

The software, which must be executed by every node of the net, has to be
unobtrusive. It has to use very few CPU and memory resources, in this way it
will be possible to run it inside low-performance computers, like Access Points,
embedded devices and old computers.

If this requirements are met, Netsukuku can be easily used to build a worldwide
distributed, anonymous and not controlled network, separated from the
Internet, without the support of any servers, ISPs or control authorities.

\section{Basic definitions}

\begin{description}
	\item[Node] We call \emph{node} any computer that is hooked up to the
		Netsukuku network.
	\item[Rnode] stands for remote node: given a node X, it is any other
		node directly linked to X, i.e. it's a neighbour of X.
	\item[Map] A map is a file, kept by each node, which contains all the
		necessary information about the network, f.e. routes and nodes
		status.
\end{description}
Example:\\
\begin{figure}[h]
	\begin{center}
		\includegraphics[scale=0.5]{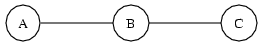}
	\end{center}
	\caption{The nodes A,B and C}
\end{figure}
A is the rnode of B.\\
B is the rnode of A and C.\\
C is the rnode of B.

\section{Network topology}
\label{sec:net_topology}

A simple topology, which doesn't impose any structure on the network, can be
memorised with a simple map. In this map, all the information regarding the
nodes of the network have to be memorised. Surely, this kind of map cannot be
utilised by Netsukuku, because it would require too much memory.
For example, even if we store just one route to reach one node and even if
this route costs one byte, we would need 1Gb of memory for a network composed
by $10^9$ nodes (the current Internet).

For this reason, it's necessary to structure the network in a convenient
topology.

\subsection{Fractal topology}
\label{sec:fractal_topology}
\subsubsection{Level 1}
First of all we'll subdivide the network in groups of 256 nodes and we'll use
the following definitions:
\begin{description}
	\item[Gnode] means group node. It is a group of nodes, i.e. a set of
		nodes. Each node of the network belongs to just one gnode.\\
		A gnode contains a maximum of 256 nodes.\\
		By writing $n \in G$ we mean that the node $n$ belongs to the
		gnode $G$.
	\item[Bnode] stands for border node. It is a node which belongs to a
		gnode G, but that is also directly linked to at least one node
		of another gnode, i.e. some of its rnodes belongs to different
		gnodes than its.\\
		By writing $b \in G$ we mean that the bnode $b$ belongs to the
		gnode $G$.
\end{description}

Example:\\
\begin{figure}[h]
	\begin{center}
		\includegraphics[scale=0.5]{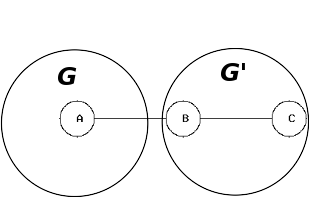}
	\end{center}
	\caption{The bnode A and B, belonging respectively to the gnode G and
	$G'$}
\end{figure}
$A \in G $, A is a node belonging to the gnode G, its rnode is B.\\
$B \in G'$, B is a node belonging to the gnode $G'$, its rnode is A.\\
A is a bnode of G, while B is a bnode of $G'$.

\subsubsection{Level n}
We further subdivide the network topology in \emph{groups of 256 groups of nodes}
and we continue to name them as gnode.\\
At this point, we repeat recursively this subdivision process until
we can group all the nodes of the network into a single gnode.

Doing so, we've structured the network in $n+1$ levels (from $0$ to $n$).\\
In the base level (level 0), there are 256 single nodes.\\
In the first level (level 1), there are 256 normal gnodes. Each of them
contains 256 single nodes.\\
In the second (level 2), 256 gnodes of level 1 forms a single \emph{group of
groups of nodes}.\\
In the third (level 3), there are 256 groups of 256 groups of 256 groups of
256 nodes.\\
Continuing in this way, we arrive at the last level (level $n$), where there
is a single group which contains the whole network.\\

The QSPN algorithm is able to operate independently on any level,
considering each gnode as a single node of level 0.
For this reason, we can view the Netsukuku topology as a fractal, where each
level is composed by single nodes.

\subsubsection*{Example}

\begin{figure}[h]
	\begin{center}
		\includegraphics[scale=0.5]{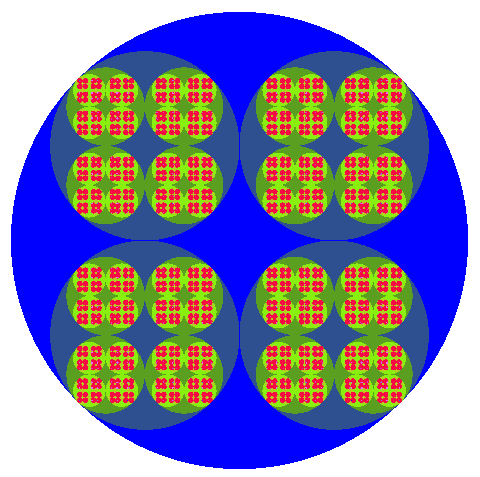}
	\end{center}
	\caption{An example of the netsukuku topology structure}
	\label{fig:fract_circle}
\end{figure}

Figure \ref{fig:fract_circle}\footnote{this figure has been taken from:
\href{http://www.ian.org/FX/Plugins.html}{http://www.ian.org/FX/Plugins.html}}
is an example of the fractal topology of Netsukuku.

In this topology, each gnode contains four nodes, i.e. each group contains
four elements. The network is structured in 6 levels.\\
The red elements, are single nodes (level 0).\\
Four nodes forms a single group of nodes (level 1).\\
A single bright green circle is a 
				  group of groups of nodes (level 2).\\
The dark green circles are        groups of groups of groups of nodes (level 3).\\
The dark blue circle are          groups of groups of groups of groups of
nodes (level 4). \\
Finally, the bright blue circle is the gnode which contains the whole network
(level 5).

\subsubsection{Membership}
Let's assign a numeric ID to each (g)gnode, starting from the last level:
\begin{enumerate}
	\item in the last level ($n$) there's only one giant gnode, thus we assign
		to it the ID ``0''. Our global ID will be:
		\[
		0
		\]
	\item in $n-1$ there are 256 gnodes, which belongs to the gnode 0 of
		level $n$, thus we assign them the IDs from $0$ to $255$.
		The global ID becomes:
		\[
		0\cdot i\quad 0\le  i\le 255
		\]
	\item we repeat the step 2 recursively gaining an ID of this form:
		\[
		0\cdot i_{n-1}\cdot i_{n-2}\cdot \dots \cdot i_0 \quad 0\le i_j\le 255,\;0\le j\le n-1
		\]
	\item since the last level is always $0$, we'll omit it and we'll
		consider only the first $n$ levels.
\end{enumerate}
In a network with a maximum of $2^{32}$ nodes (the maximum allowed by the ipv4),
there would be five levels ($n=4$), where each gnode will be composed by 256 nodes.
Therefore, the ID will be in the usual IP form:
\[
0\dots255\cdot 0\dots255\cdot 0\dots255\cdot 0\dots255
\]
For example, a single node of level 0 of the network is:
\[
3\cdot 41\cdot 5\cdot 12
\]
That said, each gnode of the network belongs to only one combination of gnodes
of the various levels. In our previous example we have:
\begin{align*}
	&g_3=3\\
	&g_2=41\\
	&g_1=5\\
	&g_0=12
\end{align*}
where each $g_i$ corresponds to the gnode ID of the level $i$. Note that $g_0$
is the ID attributed to the single node, at level 0.

\subsection{Fractal map}
The advantages of using a fractal topology are clear.\\
The node $N$, instead of memorising information about each node of the whole
network, will keep only that regarding the gnodes where it belongs to.
Suppose the node $N$ had this ID:
\[
g_3\cdot g_2\cdot g_1\cdot g_0
\]
It will store in memory information regarding:
\begin{enumerate}
	\item the 256 single nodes which belongs to its same gnode of level 1,
		or in other words, the 256 nodes of the gnode $g_1$,
	\item the 256 gnodes gnodes which belongs to its same gnode of level
		2, of in other words, the 256 gnodes of the gnode $g_2$,
	\item finally, the 256 gnodes which belongs to the gnode $g_3$.
\end{enumerate}
Note that doing so, the node $N$ will be blind to all the other gnodes. For
example, it won't know any information regarding the single nodes which belong
to all the other gnodes of level 1 different from $g_1$.\\

Even with this lack of knowledge, as we'll see later, the node $N$ is still
able to reach all the other nodes of the network.
In conclusion, $N$ only needs $256n$ entries in its map, instead of $2^{32}$. 
To clarify the ideas suppose that each entry costs one byte. In the plain
topology we needed $4Gb$, while in the fractal one we just need $256\cdot 4\;
b= 1Kb$.

\subsubsection{IP v4 and v6}
Netsukuku is both compatible with ipv4 and ipv6.\\

In ipv4 there are a maximum of $2^{32}$ IPs, thus we have five levels $n=4$.\\
In ipv6 there are a maximum of $2^{128}$ IPs, thus $n=16$.

\subsubsection{Internal and external map}
For simplicity we divide the map of the node $N$, in the \emph{internal map} and in
the \emph{external} one.  The internal map contains information regarding the
nodes belonging to $g_1$. The external map describes all the other levels of
the topology.

\subsubsection{Bnode map}
The bnode map of the node $N$  contains the information regarding the bnodes
of each level where $N$ belongs.
Some examples to clarify the ideas:\\

suppose that $N = g_3\cdot g_2\cdot g_1 \cdot g_0$
\begin{itemize}
	\item a bnode of level 0 is a single node linked with two nodes of two
		different gnodes of level 1.
	\item the bnodes of level 0, known by $N$, are only that which belong
		to the gnode $g_0$. They are all the nodes of $g_0$ which are
		linked to at least a gnode different from $g_1$.
\end{itemize}

\subsection{CIDR routing}
The QSPN, for each level, will build the routes necessary to connect each
(g)node to all the other (g)nodes of the same level. The routes will be saved
in the maps of each node.\\

If the node $N=g_3\cdot g_2\cdot g_1 \cdot g_0$ wants to reach a node $M$ which
belongs to different gnodes, f.e. $M=g_3\cdot g_2\cdot h_1 \cdot h_0$, it will
add a CIDR\cite{CIDR} route in the routing table of the kernel:\\
\emph{all the packets whose destination is $g_3\cdot g_2\cdot h_1 \cdot 0\dots
255$ will be forwarded to the gateway $X$}.\\

We'll see later how the gateway $X$ is chosen.

\section{Tracer Packets in high levels}
In the QSPN document \cite{qspndoc}, we've seen how a Tracer Packet works in a
network composed by single nodes, i.e. a gnode of level 0. \\
We'll now study its way of working on higher levels.

\subsection{A gnode is a node}
In the abstract sense a single node is an entity which:
\begin{enumerate}
	\item receives input from its links
	\item stores it in its memory
	\item computes it
	\item and sends the output of the computation over some of its links
\end{enumerate}
Thus any other entity which performs the same operations can be thought as a
single node.\\
A gnode $G$ can act as a single node too.
\begin{enumerate}
	\item A bnode $I$, which belongs to $G$, receives an input from its
		links. We call $I$ the \emph{ingress} (b)node.
	\item this input is flooded to all the nodes, of any level, of the
		gnode. The nodes will memorize the information contained in
		the input.\\
		Note that the flood is not a TP. The flooded pkt will be
		received only once by each node.
	\item A bnode $O$, of the same gnode, which is different from $I$,
		receives the flooded input and computes it.
		We call $O$ the \emph{egress} (b)node.
	\item The bnode $O$ sends the output of its computation to its
		external links, i.e. avoiding those links which connect it to
		nodes of $G$ or to the same gnode which sent the input to $I$.
\end{enumerate}

\subsubsection{Example of a wandering TP}
\begin{figure}[h]
	\begin{center}
		\includegraphics[scale=0.5]{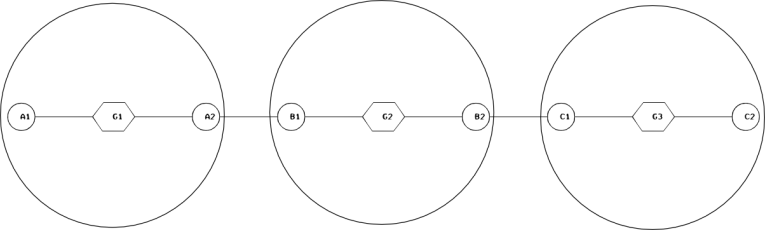}
	\end{center}
	\caption{The gnodes $G_1$, $G_2$ and $G_3$}
	\label{fig:qspn_g3}
\end{figure}

Consider the network in figure \ref{fig:qspn_g3}.\\
$G_1$, $G_2$, $G_3$ are gnodes of level 1. $A_1$ and $A_2$ belong to $G_1$. $B_1$ and $B_2$ belong to
$G_2$. $C_1$ belongs to $G_3$.\\

Suppose the node $A_1$ sends a Tracer Packet in level 1 
\footnote{Note that any node can send a TP in any level. In this case, we are
considering $A_1$, which is a bnode, to simplify the example}. 
The following will happen:
\begin{enumerate}
	\item The TP is flooded in $G_1$.
	\item $A_2$ receives the TP and appends in it the ID of the gnode
		$G_1$.
	\item $A_2$ sends the TP to $B_1$.
	\item $B_1$ receives it, updates its maps and floods it in $G_2$.
	\item All the node belonging to $G_2$ will receive the TP, updating their
		maps.
	\item This same procedure is reiterated from step 2, i.e. $B_2$ receives
		the TP, appends the ID of $G_2$ and so on until $C_1$ receives
		it.
	\item $C_1$, noticing that its gnode hasn't any links to other gnodes
		than $G_2$, will bounce back the TP to $B_2$ and at the same
		time will flood $G_3$.
	\item The TP, with the same procedure, will return back to $A_1$,
		completing the TP cycle.
\end{enumerate}

\section{QSPN v2 in high levels}
In order to use the $Q^2$ (QSPN v2) in high levels, we need to be sure that a TP,
flooded inside a lower gnode, will reach once and only once all the nodes of
the same gnode. Moreover, a good metric needs to be defined for the high
levels: what is the rtt (Round-Trip Time) and bandwidth capacity of the link
between two gnodes, how is it measured?

\subsection{Endless loops}
\begin{figure}[h]
	\begin{center}
		\includegraphics[scale=0.4]{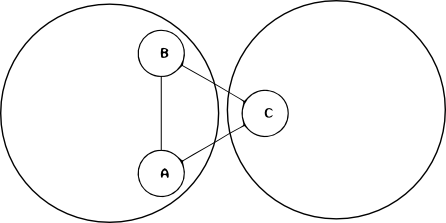}
	\end{center}
	\caption{Three bnodes, forming a cycle}
	\label{fig:3bnodes}
\end{figure}

Consider the situation in figure \ref{fig:3bnodes}.
A and B are bnodes of the node $G_1$, while $C$ is a bnode of the gnode $G_2$.
Suppose C sends a TP to B. B floods it inside $G_1$, thus $A$ receives it. At
this point, $A$ sends the packet to all its external links, and thus to $C$.
$C$ will send the packet again to $B$ and the cycle will continue.\\
\newline
This situation can be avoided if an ingress bnode, before forwarding a packet to a an
external gnode $F$, checks if the packet has already traversed $F$ itself. If
it has, the bnode won't forward it to $F$.

\subsection{Route Efficiency Measure}
We will refer to REM (Route Efficiency Measure), as a value characterising the
quality of a route. REM can be calculated in various ways, f.e. by taking in
account the total rtt and the bw capacity of the route.  We denote the REM of
a route $r$ as $REM(r)$.

\subsection{Flat levels}
\label{sec:flat}
From the point of view of the QSPN v2, the levels are ``flattened'', because
the propagation of CTPs\footnote{Continuous Tracer Packet, see \cite{qspndoc}} ignore the subdivisions of
the network in gnodes and levels. The structure of the network remains
unchanged, subdvided fractally in gnodes, but the exploration of the CTP
doesn't. A CTP doesn't consider a gnode as a single, homogenous entity. It propagates 
itself between the single nodes, tracing a precise route formed by single
hops, until it is considered interesting. 
In this way every node will have its own personal route to reach a specific gnode. 

\begin{figure}[h]
	\begin{center}
		\includegraphics[scale=0.5]{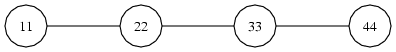}
	\end{center}
	\caption{Level 3}
	\label{figexq2}
\end{figure}

The memory limits are respected, because the CTP, once it jumped
over the border of a gnode, deletes the specific details that were useful by the nodes of
the gnode.\\
\newline
We'll know describe the rules of the Flat levels.\\
Suppose we want to explore the level $n\ge 1$ (we already know how to
explore $n=0$). Let $G_1,\dots, G_r$ be the gnode of level $n$. The generic
rule of the QSPN v2 in high levels is: 
\begin{quote}
When an ingress bnode $b$ of $G$ receives a CTP from a neighbouring bnode of
$H$, it appends in the packet its IP. The packet, called
\emph{Locked Tracer Packet} is then flooded inside $G$.
The LTP of level $n$, is a packet that is propagated with the same rules of
the CTP, i.e if it isn't interesting it's dropped, but that isn't updated
while it explores the levels inferior to $n$. When the egress bnode $b'$,
of $G$ receives the LTP, it changes it in a CTP, appends the ID of $G$ and
forwards it to all its neighbouring gnodes.
\end{quote}
Detailed rules:
\begin{enumerate}
	\item When exploring the level $n$, each inferior level must been
		already explored.\\
		To accomplish this, it sufficies that a
		node, before sending/forwarding a CTP of level $n$ has already
		sent/forwarded at least a CTP of level $n-1$. If it hasn't, it will
		queue the CTP of level $n$.
	\item Suppose that the ingress bnode $b$ belonging to a gnode $G$ of
		level $n$, has received a CTP of the same level, sent by the
		gnode $G'$. The following
		will happen: $b$ makes copy of the CTP, marking it as a LTP, appends in
		it its IP and floods it in $G$. If $b$ is also linked to a
		gnode different from $G'$, it will respect the rule
		\ref{ruleegressbnode} for egress bnodes. 
	\item \label{ruleegressbnode}
		Suppose that the egress bnode $b'$ of $G$ receives the LTP
		sent by $b$ (see above). $b'$ transforms the LTP in a CTP,
		deletes the IP added by $b$ and sets (in the TP) the REM of $G\rightarrow G'$
		as:
		\[
		REM(G\rightarrow G'):=REM(b'\rightarrow b)
		\]
		Finally, it forwards the CTP to all its external rnodes.
	\item A node $n$ of $G$, uses the received LTP to learn that $b$ is a
		bnode bordering on $G'$ and to find out the route of level $n$
		contained in the packet. It considers the REM of the route to
		reach $G'$ equal to
		\[
		REM(n\rightarrow b)
		\]
		in other words, it saves in its map that $REM(n\rightarrow
		G):=REM(n\rightarrow b)$.\\
		$n$ forwards the LTP to its rnodes without modifying it.
	\item The LTP respects the same rule of the QSPN v2: it is forwarded
		by a node only if it is carries interesting information,
		otherwise it is dropped. A
		node $n$ consider the LTP interesting if one of the following
		conditions\footnote{this isn't the complete list of
		conditions} is met:
		\begin{itemize}
			\item The LTP contains the IP of a bnode $b$, which wasn't
				already known by $n$.
			\item The LTP contains a bnode already known by $n$, but that
				borders on a new gnode $g$: with this LTP, the
				node $n$ learns that it can reach $g$ by
				passing through the bnode.
			\item The LTP contains the information of the death of
				a bnode which was used by $n$ to reach some
				gnodes.
			\item The LTP contains a new gnode, which was
				previously unknown.
			\item The LTP contains the information of the death of a
				gnode.
			\item The LTP contains an improved route to reach a
				gnode.
		\end{itemize}
		Note: the first three conditions are used by the nodes to
		build and update the bnode map.
	\item \label{rule6}
		Recall the situation of rule \ref{ruleegressbnode}.\\
		When the value $REM(b'\rightarrow b)$ changes
		considerably,
		both $b$ and $b'$ sends a CTP of level $n$ to their external
		nodes\footnote{A CTP of level $n-1$ tells $b$ and $b'$ of the
		REM change}. The CTP contains the new $REM$ value. Every node which used $G$ as hop of some routes,
		will consider the CTP interesting.\\
		This rule permits the update of high levels.
	\item \label{rule7}
		When a bnode $b$ (of any level)\footnote{a bnode $b''$, which is a
		gnode of level $l\ge 1$, loses its external link to the gnode
		$H$ of level $l+1$, when all its
		bnodes of level $l-1$ loses their link to $H$. This is a
		recursive definition.}
		loses one of its external links of
		level $n$, a CTP is sent in the level $n-1$.  The CTP contains
		the information regarding the lost link.
		The CTP is forwarded by all the (g)nodes that used the
		bnode $b$ in one of their saved routes to reach the lost link.\\
		If $b$ was the unique bnode of its gnode $G$
		bordering to gnode $G'$ of level $n$, then the link
		$G\leftrightarrow G'$ becomes broken. In this case a CTP of
		level $n$ is sent by $G'$ and by the other bnodes of
		$G$\footnote{they will know of the broken link when receiving
		the CTP of level $n-1$}.\\
		A dying bnode is equivalent to a bnode which loses all its
		external links. The only difference is that a CTP of level 0
		is sent to inform the nodes of its gnode.
\end{enumerate}
We'll now describe some rules by giving an example (see figure
\ref{figexq2}). For simplicity, we'll use the IP notation: $11.22.33$
indicates the single node $33$ belonging to the gnode $22$, belonging to the
ggnode $11$. Using the symbolic notation: $11.22.33 \in 11.22 \in 11$. We are
also assuming that the routes are symmetric.
\newline
\emph{CASE 1:  A new gnode join.}\\
We are exploring the level 3, which is formed by the gnodes 
$11,\;22,\;33,\;44$. The level 2, of each gnode, has been already
explored.
\begin{itemize}
	\item 
The gnode $44$ has just created the link to the gnode $33$. The
new physical link is maintained by the two single bnodes
$33.33.33,\;44.44.44$.
\item The bnode $33.33.33$ creates a CTP of level 3, appends in it the ID ``$33$''
and sends it to $44.44.44$.
\item The bnode $44.44.44$ receives it.
	\begin{itemize}
		\item It reflects back a new CTP to $33.33.33$, appending in it
			the ID ``$44$''.
		\item It appends in the received CTP, its IP, marks it as a
			LTP and sends it to its gnode. The LTP contains the
			path $33\rightarrow 44.44.44$.
			\begin{itemize}
				\item The node $44.44.x$, receiving the LTP,
					learns the following: $44.44$ is a bnode of
					level 2 that can be used to reach
					$33$, and $44.44.44$ is a bnode of its
					own gnode which borders on $33$. It
					also learns its own REM for the
					route to $33$: \[REM(44.44.x
					\rightarrow
					33):=REM(44.44.x\rightarrow 44.44.44)\]
					$44.44.x$ continues to forward the LTP
					(without modifying it).
				\item The LTP reaches the node $44.y.z$,
					node of $44.y$. $44.y.z$, from the LTP
					learns that: $44.44$ is a bnode of
                                        level 2 that can be used to reach
					$33$. Since the level $2$ has been
					already explored, $44.y.z$ already
					knows a route to reach $44.44$.
				\item The LTP having reached all the nodes of
					$44$ ceases to being forwarded.
			\end{itemize}
	\end{itemize}
\item The bnode $33.33.33$ receives the reflected CTP sent by $44.44.44$.
	\begin{itemize}
		\item It appends in it its IP, marks it as a LTP and sends it to its gnode.
			\begin{itemize}
				\item The LTP is propagated in $33$ (in the same way of the
					previous LTP).
			\end{itemize}
	\end{itemize}
\item The bnode $33.r.s$, which borders on $22$, receives the LTP.
	\begin{itemize}
		\item It modifies it in a CTP: it removes the IP $33.33.33$
			and appends the ID $33$. The CTP is now $44\rightarrow
			33$. Finally, it sets the first REM of the CTP:
			\[ REM_{CTP}(33 \leftrightarrow 44):= REM(33.r.s
			\leftrightarrow 33.33)\;\;\;\textrm{ see note }
			\footnote{The node $33.r.s$
			knows that $33.33$ is a bnode of level 2 used to reach
			$44$ because it is written in the LTP. Moreover it
			knows a route to reach $33.33$, the level $2$ has been
			already explored. Thus $REM(33.r.s\leftrightarrow
			33.33)$ is a known measure.}\]
			The CTP is then sent to $22.22.22$, bnode of $22$.
	\end{itemize}
\item The bnode $22.22.22$ receives the CTP. It sends a LTP in $22$.
\item The bnode $22.t.u$, which borders on $11$, receives the LTP. It 
	appends $22$ in it and sends it to $11$. The CTP is now: $44
	\rightarrow 33\rightarrow 22$.
\item A bnode of $11$ receives it, and reflect it to $22$. It sends also the
	LTP to $11$.
\item \dots
\item The reflected CTP is finally received by the node $44.44.44$, which
	sends the LTP to $44$ and reflects the CTP to $33.33.33$. The bnode
	$33.33.33$, drops immediately the CTP, because it doesn't contain any
	interesting information.
\item the exploration is completed.
\end{itemize}

%
%

\subsection{Unique flood}
Suppose that a TP has been flooded inside the gnode $G$.
Suppose also that the node $n \in G$ receives two duplicate packets.\\
The $Q^2$ instructs the node $n$ to keep forwarding only the interesting packets.
A packet, which is a perfect copy of a packet already received, is always
uninteresting. Therefore, the node $n$ will drop the second copy of the
received TP. 

However, in this case the node $n$ should not consider the REM
values saved in the TP, because two TPs, which have crossed the same
gnodes in the same order, might have different REM values.
\\

In conclusion, whenever two or more TPs, which have crossed the same high level
route, are received by a node of $G$, only one of them will be forwarded. In
this way, the successive nodes will receive only one copy of the same packet.

\section{Network dynamics}
When a part of the network changes considerably, the maps of the involved
levels must be updated.

\subsection{Radar}
Every node has its own radar, which periodically sends a broadcast request to
all its (physical) near nodes. By collecting the replies, the radar is able to
determine the active rnodes of the node and the quality of its links.

\subsection{Level 0}
\label{sec:netdyn-level0}
In level 0, a CTP is sent every time a node joins the network, dies or every
time the change of the quality of a link exceeds a predefined delta value. The
CTP is also is restricted to the gnode where it has been originated: a
bnode won't send it to external nodes.\\
\newline
When a node joins, it won't send a CTP, only its rnodes will.
Their CTP will be directed to the node. More formally:\\
if the node $n$ joins, that is if the node $n$ creates a link
with the nodes $r_1,\dots,r_n$, with $n\ge 1$, then a CTP will
be sent by each rnode $r_i$ \emph{only} to $n$. This means that if $r_j$ is connected to a node $s$, when
$n$ joins, $r_j$ will send the CTP only to $n$ and not to
$s$.\\
This saves the propagation of two CTP in the following situation:
\[\dots \leftrightarrow A\leftrightarrow B\leftrightarrow C\leftrightarrow\dots\]
suppose that the node $B$ joins. As a consequence of the rule we've described,
the node $C$ will send a CTP to $B$ and $A$ to $B$, thus only two distinct CTP are
generated. Their path is $A\rightarrow B\rightarrow C\rightarrow \dots$ and
the reverse. If instead the node $B$ sends a CTP to $C$ and then $A$ to $B$,
two distinct CTP will explore the same verse: $B\rightarrow
C\rightarrow\dots $ and $A\rightarrow B\rightarrow C\rightarrow \dots$, the
same is for the reverse, thus, in total, four distinct CTP are propagated.\\
\newline
Let $A\stackrel{l}{\leftrightarrow}B$ be a link. If its quality changes $B$
will send a CTP to $A$ and vice-versa.\\
\newline
When a node dies, all its rnodes will send a CTP to all their rnodes. The CTP
will include, as the first hop, the ID of the dead node, with a flag which
indicates its death. For example:
\[A\leftrightarrow B\leftrightarrow C\leftrightarrow\dots\]
if $A$ dies, then $B$ will send the following CTP to $C$:
$\stackrel{+}{A}\;\rightarrow B$\\
\newline
In order to prevent false positives, the nodes won't immediately send the CTP, but
will wait a small amount of time. Only if the change persists, they will send it.

\subsection{Level n}
\label{sec:netdyn-leveln}
The dynamics for the update of high levels are mainly governed by rule
\ref{rule6} and \ref{rule7} of flat levels (see \ref{sec:flat}). In this
paragraph, we'll specify some details.\\
\newline
Rule \ref{rule6} says: ``when the value $REM(b'\rightarrow b)$ changes
considerably, a new CTP is sent''.
By considerably we mean that the difference of the current REM value with the
previous one exceedes a predefined threshold. More precisely: let
$A\rightarrow B$ be a route of level $n$. If \[|REM_{\textrm{now}}(A\rightarrow B)-
REM_{prev}(A\rightarrow B)| > \Delta_n\]
then the route has changed considerably. $\Delta_n$ is proportional to $n$, in this way the nodes will be
more sensible to minor changes of lower levels and they will consider only
relevant changes of higher levels.\\
\newline
Also in high levels, a bnode won't immediately send a new CTP, but it
will delay it for a small amount of time.

\subsection{Hooking phase}
A new node joins the network when it has been able to create at least one
physical link to an active Netsukuku node and when it has correctly executed
the hooking procedure. In this paragraph, we describe loosely the hooking
phase of a new node.

Suppose that the node $n$ has established a physical link to at least one Netsukuku
node. In order to become an active Netsukuku node, $n$ has to \emph{hook} to
its rnodes.

During the hook, $n$ will exchange vital information with its rnodes,
it will choose its new IP and it will finally become part of a gnode.

The hook procedure is formed by these general steps:
\begin{enumerate}
	\item The node $n$ chooses an IP in the range of $10.0.0.1 \le IP \le
		10.0.0.255$.
	\item It launches the first radar to see what its rnodes are. If not a
		single node is found, it creates a new gnode and ends hooking
		phase.
	\item At this point, $n$ asks to its nearer rnode the list of all the
		available free nodes presents inside the gnode of the rnode.\\
		If the rnode rejects the request (the gnode might be full),
		the node $n$ contacts another rnode.
	\item $n$ chooses an IP from the received list of free nodes and sets
		it on its network interface.
	\item $n$ will then download the external map from the same rnode.
		Looking at the external map, it will be able to determine if
		it has to create a new gnode. If it has, it creates it and
		ends the hooking.
	\item $n$ gets the internal and the bnode map from the same rnode.
	\item $n$ launches a second radar and updates its routing table.
	\item All the rnodes of $n$ send a CTP to update the maps.
\end{enumerate}

\subsection{Gnode hook}
When a node creates a new gnode, it will choose a random gnode ID, and thus
a random ip.\\
Suppose that two isolated gnodes get the same gnode ID. When they will be
linked, they'll enter in conflict.\\
The solution to this problem is to let each new gnode hook as a normal node
would. You can find more information about this in the NTK\_RFC 001\cite{gnodecontiguity}.


%

\section{ChangeLog}
\begin{itemize}
	\item \verb|March 2007|
		\begin{itemize}
			\item Description of the Flat levels (sec. \ref{sec:flat})
			\item Section \ref{sec:netdyn-level0} ``Network dynamics - Level 0'' expanded.
			\item Section \ref{sec:netdyn-leveln} ``Network
				dynamics - Level n'' updated: the references to
				the pre-Flatlevels REM metric have been
				removed.
		\end{itemize}
	\item \verb|October 2006|\\
		Initial release.
\end{itemize}


\newpage

\begin{center}
\verb|^_^|
\end{center}
\end{document}